\documentclass[aps,pra,reprint]{revtex4-1}
\usepackage{amsmath}
\usepackage{amssymb}
\usepackage{graphicx}
\usepackage{epstopdf}
\usepackage{esint}
\usepackage{tikz}
\newcommand{\itext}[1]{\mbox{\footnotesize{#1}}}

\begin{document}

\title{Modal theory of modified spontaneous emission for a hybrid plasmonic photonic-crystal cavity system}

\author{Mohsen Kamandar Dezfouli$^{\mathrm{1}}$, Reuven Gordon$^{\mathrm{2}}$, and Stephen Hughes$^{\mathrm{1}}$}

\affiliation{$^{\mathrm{1}}$Department of Physics, Engineering Physics and Astronomy, Queen's
University, Kingston, ON K7L 3N6, Canada \\
$^{\mathrm{2}}$Department of Electrical and Computer Engineering, University of Victoria, Victoria, BC V8W 3P2, Canada}

\begin{abstract}{We present an analytical modal description of the rich physics involved in hybrid plasmonic-photonic devices that is confirmed by full dipole solutions of Maxwell's equations. Strong frequency-dependence for the spontaneous emission decay rate of a quantum dipole emitter coupled to these hybrid structures is predicted. In particular, it is shown that the Fano-type resonances reported experimentally in hybrid plasmonic systems, arise from a very large interference between dominant quasinormal modes of the systems in the frequency range of interest. The presented model forms an efficient  basis for modelling quantum light-matter interactions in these complex hybrid systems and also enables the quantitativ prediction and understanding of non-radiative coupling losses.}
\end{abstract}

\maketitle

Plasmonic devices show great promise for applications in quantum photonics and sensing technologies \cite{Kneipp1997,Willets2007,Zhang2013}, due in part to the strong local field confinement below the diffraction limit. However, metals naturally have Ohmic losses and one must characterize both enhancement effects and Ohmic dissipation on an equal footing. On the other hand, recently, gold nanorod dimer structures have been shown to pose good single photon output $\beta$-factors {(i.e., the fraction of the dipole-emitted radiated power that is available in the far field)}   of around 60\% \cite{rgche2014ol}, as well as very strong spontaneous emission (SE) rate enhancements  \cite{russell2012,vesseur2010,rgche2014ol}. In addition, integration of plasmonic structures with photonic crystal (PC) cavity platforms has been shown to offer new possibilities \cite{Barth2010,Eter2014} that can benefit both from higher  $Q$ of the PC sub-system and the stronger field enhancements and tighter light confinement by the plasmonic sub-system. In particular, plasmonic devices offer an extremely wide bandwidth compare to dielectric devices such as PC cavities, because they have intrinsically low  quality ($Q$) factors. Therefore,  by coupling these two systems together, the possibility of introducing very fine spectral features (due to PC-cavity) within a  broad operating band (due to plasmonic system) can be investigated. However, the theoretical description of such hybrid devices is rather scarce and  particularly complicated because the traditional single mode cavity model fails for a number of reasons. Moreover, the rich physics behind Fano-type resonances that have been seen in hybrid plasmonic cavity systems \citep{Barth2010} is also  not yet well understood or explained.

\begin{figure}[ht!]
\begin{tikzpicture}
\node[anchor=south west,inner sep=0] at (0,0) {\includegraphics[width=\columnwidth]{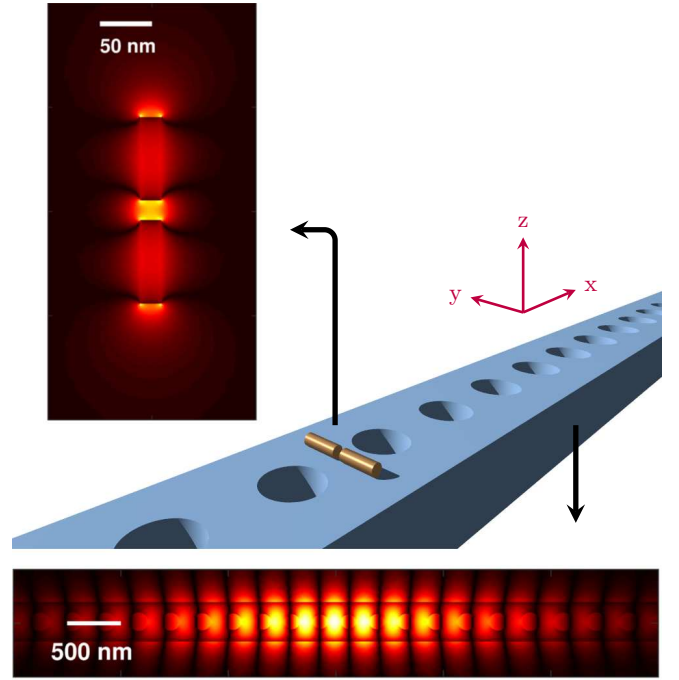}};
\draw[line width=2,rounded corners,-stealth] (4.3,3.4) -- (4.3,6) -- (3.7,6);
\draw[line width=2,rounded corners,-stealth] (7.5,3.4) -- (7.5,2.1);
\begin{scope}[shift={(0.8,0.4)},color=purple]
\draw[line width=1,-stealth] (6,4.5) -- (6.7,4.8);
\node at (6.9,4.85) {x};
\draw[line width=1,-stealth] (6,4.5) -- (5.3,4.7);
\node at (5.1,4.7) {y};
\draw[line width=1,-stealth] (6,4.5) -- (6,5.5);
\node at (6,5.7) {z};
\end{scope}
\end{tikzpicture}
\caption{Schematic of the hybrid device where a gold dimer of nanorods is placed on top of a nanobeam PC-cavity (see inside text for design parameters). The top colormap shows the $|E_y|^2$ of the dimer-only QNM when placed in free space in the middle of dimer. The colormap on the bottom, shows the $|E_y|^2$ of the PC beam-only QNM in the middle of the slab. The QNM frequencies for the dimer and the PC-cavity are $\tilde{\omega}_{\rm di}({\rm eV})=1.7803 - 0.0678i$ and $\tilde{\omega}_{\rm pc}({\rm eV})=1.6156 - 2.4539\times10^{-6}i$, respectively. The origin of our coordinate system, $(x,y,z)=(0,0,0)$, is placed exactly in the middle of the dimer gap.\label{fig:device}}
\end{figure}

In this Letter, we introduce a system that consists of a gold dimer \cite{rgche2014ol} placed on top of a nanobeam PC-cavity \cite{Mukherjee2011}, and show how one can achieve very strong manipulation of SE decay rate. Perturbative analysis of similar structures has been done in the past in order to estimate the frequency shift of the system resonances due to the interplay between the sub-systems \cite{Mukherjee2012}, however, full analytic characterization of the hybrid system using the system fundamental modes remain challenging. Using a rigorous modal description, we show that when a dipole is placed in between the dimer, then a  very large degree of interference between the quasinormal modes (QNMs) \cite{Philip2014} of the two cavity modes in the hybrid system takes place. The QNMs are the frequency domain mode
solutions to the wave equation with open boundary conditions (the Silver-M\"uller radiation condition) \cite{Leung1994}. Using this modal description, we study how the Purcell factor and $\beta$ factor changes as a function of frequency and dipole position and verify the accuracy of our approach by comparing against full dipole solutions to Maxwell equations. This modal description not only explains all the underlying physics of optical enhancement and quenching, but it can also be used as a foundation of studying quantum plasmonics in these hybrid systems, where the use of a Purcell factor and modal theory have been questioned \cite{Koenderink2010}.

A schematic of the proposed device is shown in Fig.~\ref{fig:device}, where two colormaps show the spatial profile of the main QNM of the individual sub-systems, namely the gold dimer (top) and the nanobeam PC-cavity (bottom). The dominant response of the combined system is driven by the two hybridized QNMs that are partly dimer-like and partly PC-like. Since the two QNMs of interest do strongly overlap in frequency, we use a frequency domain technique  \cite{Bai2013} to calculate them, using the commercial software COMSOL. We then use the two obtained QNMs as a basis to model the response of the system and confirm this prediction with full dipole calculations. We first start with description of the theoretical formalism appropriate for study of such hybrid devices in terms of the QNMs of the system.

Quasinormal modes, $\mathbf{\tilde f}_{\mu}\left(\mathbf{r}\right)$, are solutions to a non-Hermition Maxwell's problem  that are associated with a complex eigenfrequency $\tilde{\omega}_{\mu}=\omega_\mu-i\gamma_\mu$, the imaginary part of which is a measure of energy leakage in the system, quantified by the quality factor $Q_\mu=\omega_\mu/2\gamma$. The system response can be analyzed in the basis of the QNMs through the Greens function expansion \cite{Lee1999}:
\begin{equation}
\mathbf{G}^{\rm QNM}\left(\mathbf{r}_1,\mathbf{r}_2;\omega\right)=\sum_{\mu}\frac{\omega^2}{2\tilde{\omega}_{\mu}\left(\tilde{\omega}_{\mu}-\omega\right)}\,\mathbf{\tilde f}_{\mu}\left(\mathbf{r}_1\right)\mathbf{\tilde f}_{\mu}\left(\mathbf{r}_2\right),
\label{GF}
\end{equation}
which can then be used to obtain the SE decay rate of a quantum dipole emitter, $\mathbf{d}={\rm d}{\hat y}$, placed at $\mathbf{r}_0=(0,0,0)$ (see Fig. \ref{fig:device}) and oriented along the $y$ axis, through
\begin{equation}
\Gamma=\frac{2}{\hbar\epsilon_0}\,\mathbf{d}\cdot{\rm Im}\{\mathbf{G}^{\rm QNM}\left(\mathbf{r}_0,\mathbf{r}_0;\omega\right)\}\cdot\mathbf{d}.
\label{PF}
\end{equation}
Using the Green function for the homogeneous medium, $\mathbf{G}^{\rm B}$, in \eqref{PF}, one can calculate the free space SE decay rate, $\Gamma_0$, and therefore the projected SE enhancement factor along dipole direction $F_y=\Gamma/\Gamma_0$.
Note that the QNMs can be directly obtained in normalized form \cite{Bai2013} by using  Eq.~(\ref{GF}) and obtaining the scattered field solution to a point dipole; the corresponding complex mode-volume for use in Purcell's formula is then defined from ${\rm V^{eff}}=1/[n^2_b\,\tilde{\mathbf{f}}^2_{\mu}\left(\mathbf{r}_0\right)]$  \cite{Kristensen2012,Sauvan2013}.

In addition, the QNMs of the system can be also used to describe the non-radiative decay rate of the dipole from \cite{Novotney2006}
\begin{equation}
\Gamma_{\rm NR}=\frac{2}{\hbar\omega\epsilon_0}\int_{\rm V} {\rm Re}\{\mathbf{j}(\mathbf{r})\cdot\mathbf{E}^{*}(\mathbf{r})\}{\rm d}^3\mathbf{r},
\label{NR}
\end{equation}
where $\mathbf{E}(\mathbf{r})=\mathbf{G}^{\rm QNM}\left(\mathbf{r},\mathbf{r}_0;\omega\right)\cdot\mathbf{d}$ is the field emitted by the dipole at ${\bf r}_0$ and $\mathbf{j}(\mathbf{r})=\epsilon_0\,\omega\,{\rm Im}\{\epsilon(\mathbf{r})\}\mathbf{E}(\mathbf{r})$ is the current density induced by the dipole over the metal volume, V. Accordingly, the ratio between the radiated power to the far field and the total radiated power by the dipole is $\beta=1-\Gamma_{\rm NR}/\Gamma$.

\begin{figure}
\includegraphics[width=\columnwidth]{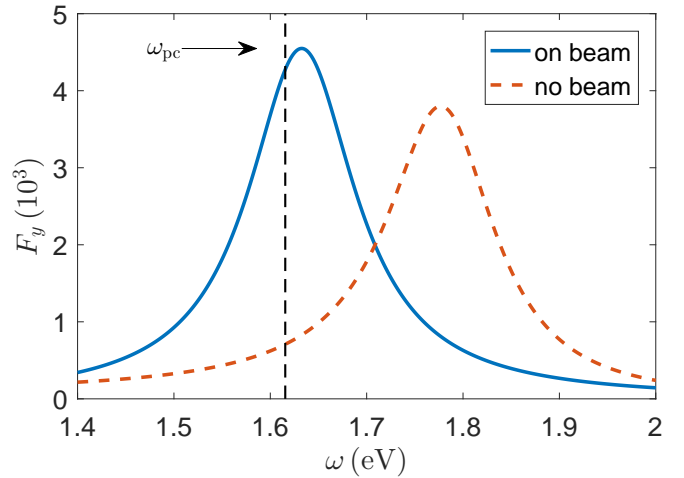}
\caption{Comparison between the Purcell factor, $F_y$, of a dipole placed in between gold nanorods, when placed in free space (solid blue) and when placed  on top of the dielectric beam without  any PC holes (dashed red). In both cases, the dipole is placed in between the gold nanorods (center position) and is directed along the longitudinal axis of the rods.\label{fig:dimer-pf}}
\end{figure}

The gold dimer is made of two nanorods with radius and rod height of $r_{\itext{Au}}=10\,{\rm nm}$ and $h_{\itext{Au}}=80\,{\rm nm}$, respectively. The dimer is  in free space with background refractive index of $n_{b}=1$. The Lorentz plasmon model, $\epsilon(\omega)=1-\omega_p^2/(\omega(\omega+i\gamma_p))$, where the plasma frequency of $\omega_p = 1.26\times10^{16}\,{\rm rads/s}$ and collision rate of $\gamma_p = 1.41\times10^{14}\,{\rm rads/s}$ is used to describe the frequency response of the gold in our system. This dimer has a single mode behavior over a wide range of frequencies as shown in Fig.~\ref{fig:dimer-pf}. The dimer alone has a very large $F_y=3800$ corresponding to the mode profile shown in Fig.~\ref{fig:device}. If one brings the gold dimer close to the surface of a nanobeam without any PC structure patterned (i.e., a beam without holes), there will be two main effects:  the resonance frequency of the dimer is red shifted and  the decay rate becomes enhanced further. This is shown in the same figure for comparison, when the gold dimer is placed 5 nm away from the surface of the beam. The accurate knowledge of the red-shifting is  important in obtaining good coupling between the two sub-systems, as investigated here.

The nanobeam PC-cavity is made of silicon-nitride with refractive
index of $n=2.04$ \cite{Bai2013}. The height and the width of beam are $h=200\,nm$
and $w=367\,nm$, respectively. Following \cite{Loncar2008}, the nanobeam design includes two sections, namely mirror and taper, where the hole radius and spacing are different. The taper section is made of 7 holes, such that their radius were decreased form 86 nm to 68 nm and their spacing was decreased from 306 nm to 264 nm, in a linear fashion. On the ends of the taper section, the mirror section is designed such that 17 holes of fixed radius $r=86\,{\rm nm}$ with fixed spacing of $a=306\,{\rm nm}$ are used. The length of the cavity region in between the two smallest holes, in the very middle of the structure, is chosen to be 126 nm. This design supports one main QNM of interest at $\tilde{\omega}_c({\rm eV})=1.6153 - 2.5928\times10^{-6}i$,
corresponding to a large quality factor of $Q=3\times 10^{5}$. The mode profile for this QNM is shown in Fig.\ref{fig:device}, as well. Our investigations show that this PC-cavity supports additional QNMs at lower frequencies with lower Qs that can be effectively ignored for working in the frequency range of our interest;  this will be made more clear when discussing the $F_y$ characteristics of the device below.

\begin{figure}
\includegraphics[width=\columnwidth]{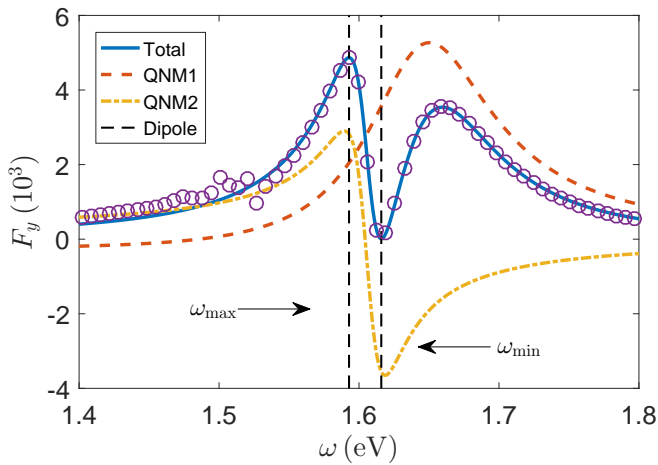}
\caption{Generalized Purcell factor, $P_{F}$ calculated for the hybrid structure where the $y$-polarized excitation
dipole is placed at $\mathbf{r}_0$. The solid blue
line is the analytic calculation using an expansion of the dominant QNMs of the system, while the purple circles are the full dipole calculations. Small disagreement at lower frequencies comes from other QNMs of the PC-cavity
that are not included in the  analytical study.\label{fig:hybrid-pf}}
\end{figure}

When the gold dimer is placed on top of the nanobeam  cavity, the two individual modes discussed above become strongly hybridized. The resonance frequencies of the two hybridized QNMs are found to be
$\tilde{\omega}_{1}({\rm eV}) = 1.6429 - 0.0548i$ and 
$\tilde{\omega}_{2}({\rm eV}) = 1.6063 - 0.0144i$, corresponding to $Q_{1}=15$ and $Q_{2}=55$, respectively. The associated generalized mode volumes at $\mathbf{r}_0$ are also estimated to be $V^{\rm eff}_1({\lambda_1}^3)=(1.96+0.68i) \times 10^{-4}$ and $V^{\rm eff}_2({\lambda_2}^3)=(-0.86-6.39i) \times 10^{-4}$.
Note that, the latter is indeed negative which originates from the interference between two sub-systems and can be understood by looking at the actual contributions from each QNM to the total decay rate of the dipole (see Fig.\ref{fig:hybrid-pf}); being negative suggests that this does volume does not represent a physical volume, but rather is a quantity with dimensions of volume that is required for use in the calculation of the Purcell factor. These two coupled QNMs are believed to be responsible for the dominant response of the system over the frequency range of our interest. In order to confirm this and represent the accuracy of the analytical model in the basis of QNMs, the predicted $F_y$ is compared with the full dipole calculations at different frequencies. In Fig.~\ref{fig:hybrid-pf}, the $F_y$ is plotted, both using full dipole calculations in circles and analytic calculations in blue solid line, where an excellent agreement between the two is obtained. In the same figure, we have also plotted the contributions from each individual mode. Note that, each of the individual enhancement factors do not necessarily represent physically meaningful quantities, however, the total enhancement (i.e., simply adding individual contributions) is confirmed to be all positive, and well behaved. There are also small oscillations present in the full dipole calculations at lower frequencies that are not captured by the analytic model that uses only a two mode expansion. As mentioned before, these are due to contributions form lower frequency modes of the PC-cavity and can safely ignored. The modal description provides the full system response including the enhanced decay rate of a given dipole at any location over a wide range of frequencies, where normally one single dipole calculation must be performed per frequency point per spatial position. In addition, as will be discussed next, the modal expansion brings insight into the underlying physics that is not normally available from full dipole calculations. Moreover, the obtained Green function of the system can then be used to explore  quantum dynamics of dipole emitters coupled to this system \cite{rchge2015}.

\begin{figure}
\centering
\includegraphics[width=\columnwidth]{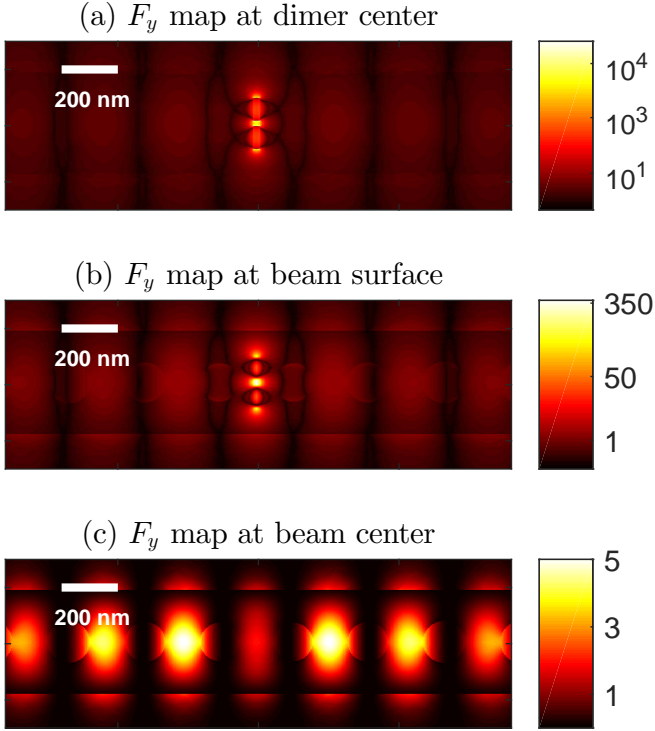}
\caption{(a)-(c) show $F_y$ at $z=0$ (in between the dimer gap), $z=-15\,{\rm nm}$ (at the beam surface) and $z=-115\,{\rm nm}$ (in the middle of the beam), respectively plane. The field is stronger around the gold dimer, however, for the low-Q
mode it is even stronger compare to the high-Q mode. Since the field intensity around dimer is orders of magnitude larger than in PC beam, a nonlinear scale is used to bring out the mode features.\label{fig:modes}}
\end{figure}

The maximum modal SE enhancement is  approximately $F_y=4900$ which is similar to that of achieved for the dimer on top of the slab alone. However, this maximum now occurs at a different frequency closer to the frequency of higher-Q QNM which itself is close the resonance of the bare PC-cavity. The accurate knowledge of the shifting, that is a consequence  of an effective coupling between the two components of the hybrid device, is well beyond weak coupling. But more importantly, just next to the maximum, as a result of the very strong interference between the two individual QNMs, a minimum enhancement occurs that is drastically different in magnitude compare to the maximum; indeed the $F_y$ is almost reduced to zero (approximately to $F_y=4$), which can be utilized as a switching mechanism between the two fundamentally different response regimes of the system. For example, a quantum dot placed in the dimer gap will become extremely dark if excited at this particular frequency, whereas in contrast becomes quickly bright when one moves away from this minimum point.

As mentioned earlier, hybrid QNMs of the system inherit features form both dimer and PC-cavity. However, the low-Q mode is more dimer-like than the high-Q mode. Indeed, for the particular structure under study, the magnitude of the field in between dimer gap is found to be more than an order of magnitude stronger than in the center of the nanobeam, for both of the QNMs, which is another indication of significant hybridization of the individual QNMs of the system. To help quantify  the hybrid characteristics of these modes it is useful to look at a spatial map of the enhancement factor, $F_y$. In Fig.~\ref{fig:modes}, we \ plot $F_y$ at three different heights ($z$ values) over a rectangular $xy$ cut: (a) at the center of the dimer gold, (b) on the surface of the nanobeam and (c) at the nanobeam center. Note that, the system response behaves mostly dimer-like in Fig.~\ref{fig:modes}.(a) whereas in contrast becomes more PC-cavity like in Fig.~\ref{fig:modes}.(c). However, at any height both components contribute to the response. These maps are calculated at the frequency of the maximum enhancement factor in Fig.~\ref{fig:hybrid-pf}, shown by the dashed line. A drastic decrease in the $F_y$ is seen when moving away from dimer and closer to the nanobeam cavity, as shown in Fig.~\ref{fig:hybrid-pf}. However, this trend
is not always obtained and depends on frequency, e.g., at the exact frequency that the minimum takes place, we found that quite the opposite occurs and $F_y$ will increase from its minimum value in the dimer gap to higher values in the middle of the nanobeam. This is a non-trivial feature of this hybrid device that originates from significant hybridization discussed earlier. It should be also noted that, because the dimer greatly shapes the structures of both of the QNMs in this device, the increase in the $F_y$ mentioned later is not as drastic as the decrease mentioned in the previous scenario.

\begin{figure}
\includegraphics[width=\columnwidth]{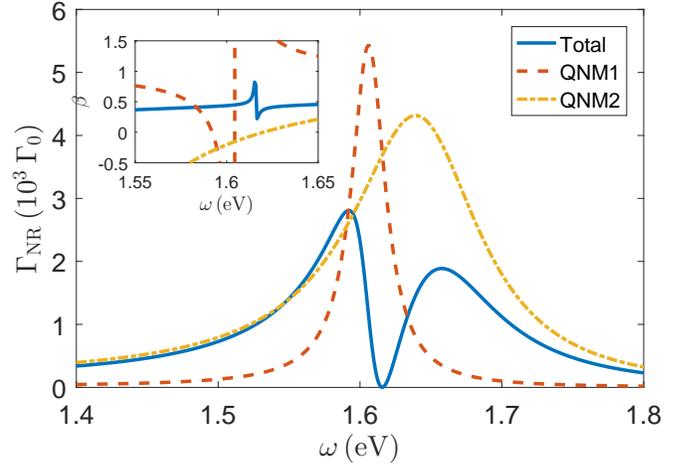}
\caption{$\Gamma_{\rm NR}$ calculated for the hybrid structure where the excitation
dipole is placed at $\mathbf{r}_0$. The solid blue
line is the analytic calculation using expansion of the system response in terms of dominant QNMs of the system, while the purple circles are the full dipole calculations. Disagreement at lower frequencies comes from other QNMs of the PC-cavity that are not included in the present analytical study.\label{fig:hybrid-gama}}
\end{figure}

As for any plasmonic device, another  important quantity of interest is the non-radiative decay rate of \eqref{NR}, that can be quantified analytically using the system QNMs. Dealing with \eqref{NR} is trivial in the case of a single mode system,  however extra caution must be taken when there are two (or more) QNMs involved. In the present case, the field generated over the lossy region has two dominant contributions through different QNMs, so\begin{equation}
\mathbf{E}(\mathbf{r})=\frac{\omega^2\mathbf{\tilde f}_1(\mathbf{r}_0)\cdot\mathbf{d}}{2\tilde{\omega_1}\left(\tilde{\omega}_1-\omega\right)}\,\mathbf{\tilde f}_1(\mathbf{r})+\frac{\omega^2\mathbf{\tilde f}_2(\mathbf{r}_0)\cdot\mathbf{d}}{2\tilde{\omega_2}\left(\tilde{\omega}_2-\omega\right)}\,\mathbf{\tilde f}_2(\mathbf{r}).
\label{NR}
\end{equation}
Therefore, there will be cross-coupling of the two QNMs of the hybrid system to be integrated over the metallic region.  In contrast to the total decay rate, $F_y$, the simple adding of the contributions from single QNMs do not add up to the total $\Gamma_{\rm NR}$. In Fig. \ref{fig:hybrid-gama}, we plot the $\Gamma_{\rm NR}$ for the hybrid system where the general trend of the total non-radiative decay rate is similar to the total decay rate. In the same figure, we have also plotted the pure contributions to the non-radiative decay rate form each individual QNM, which again, these do not represent physically meaningful quantities. The inset in Fig. \ref{fig:hybrid-gama} shows the $\beta$-factor, where we find a relatively fixed value of approximately $\beta=0.46$ over wide range of frequencies, except the unusual discontinuity that occurs exactly at the frequency where the minimum $F_y$ occurs. Higher values for the $\beta$-factor can be achieved if moving slightly away from this minimum point to higher frequencies, but note that practically very low total decay rates are involved.  

In conclusion, we have introduced a hybrid plasmonic-PC system that is capable of very strong modification of the SE decay rate of dipole emitters when placed right in the middle of the dimer gap. The drastic change from $F_y=4900$ to $F_y=4$ can be utilized as a switching knob to trigger into fundamentally different response regimes of this system. In addition, the overall $\beta$-factor of about 0.46 is achieved that suggest reasonable out coupling of the light to the far field. {To study the system, we have used a modal description that is capable of drawing a clear picture of the physics behind its non-trivial response. With the efficient modal description provided here, the full decay rate characteristics of the system are available, which can also be used to study quantum light-matter interaction.

\section*{Funding}
This work was supported by Queen's University and the Natural Sciences and Engineering Research Council of Canada.

\bibliographystyle{apsrev4-1}
\bibliography{text}

\end{document}